# Qualitative Approaches to Voice UX

**Katie Seaborn**

*Tokyo Institute of Technology*
*Tokyo, Japan*

**Jacqueline Urakami**

*Tokyo Institute of Technology*
*Tokyo, Japan*

**Peter Pennefather**

*gDial Inc.*
*Toronto, Canada*

**Norihisa P. Miyake**

*RIKEN Center for Advanced Intelligence Project (AIP)*
*Tokyo, Japan*

**ABSTRACT:** Voice is a natural mode of expression offered by modern computer-based systems. Qualitative perspectives on voice-based user experiences (voice UX) offer rich descriptions of complex interactions that numbers alone cannot fully represent. We conducted a systematic review of the literature on qualitative approaches to voice UX, capturing the nature of this body of work in a systematic map and offering a qualitative synthesis of findings. We highlight the benefits of qualitative methods for voice UX research, identify opportunities for increasing rigour in methods and outcomes, and distill patterns of experience across a diversity of devices and modes of qualitative praxis.

## 1  Introduction

Voice is ubiquitous in human communication, expression, and embodiment [79,85]. This has inspired vested efforts in crafting technologies and conducting research on voice interaction, speech recognition, and vocal synthesis in computer systems. Modern handheld devices and virtual assistants demonstrate an increasing prevalence of, or at least a desire for, two-way voice-based interaction [17,24,139]. Global sales of such agents and devices as well as smart speakers indicate commercial and consumer interest. For instance, Amazon shipped 40.1 million smart speaker units in 2021 alone, with 60% of the market share being younger people aged 18-24[1]. Within research and industrial R&D spaces, decades of work has pushed for advances in voice-enabled speech interfaces and environments. Notable examples include work on natural language processing (NLP) in machine learning (ML) fields [22]; efforts to replicate realistic human-like vocals in text-to-speech (TTS) systems [16]; and a move towards natural forms of expression during voice and speech interaction, perhaps best represented by embodied conversational agents (ECAs) [19] and conversational user interfaces (CUIs) [102]. Application areas range from interfaces for Internet of Things (IoT) systems, smart cars that speak, accessibility for blind and low-vision users, and also hands-free contexts of use [26,139]. Suffice to say, "visions" of voice interaction, advances in technology, and ongoing commercial viability have drawn the attention of researchers across a range of technical fields of study to the question of what voice user experiences (UX) modern forms of voice interaction in computer-based systems can, and should, provide.

Exploring voice UX, indeed any form of UX, requires multiple forms of inquiry. UX as a human factor and research construct is not easily reduced to a series of variables or best represented by numbers [55,56,82,86]. Experience is by definition qualitative, subjective, impressionistic, and encompassing. It can be difficult to describe, complex, and can even shift over time for one individual. This is not to say that quantitative approaches to voice UX are invalid. Indeed, previous literature surveys and systematic reviews [4,17,24,139,141,153] have fruitfully mapped out various modes of quantitative inquiry, measures and instruments, and generalized (or generalizable) findings on the UX that computer voice, speech interaction, and conversational formats can offer. What is needed, as pointed out in the limitations of recent reviews [24,139], is a greater understanding of the *qualitative* side. This means not only the qualities of experience that people can have with voice and speech in computer systems, but also what qualitative and mixed methodologies have been most rewarding, all towards generating new questions, understanding quantitative results, and setting the stage for theory development [32]. This may be especially important for those of us in computer science and engineering [78] who may not be exposed to these forms of inquiry, understand their purpose, or recognize their value.

---

[1] https://www.statista.com/topics/4748/smart-speakers/

To this end, we have conducted a systematic review of the literature on qualitative approaches to voice UX. Our objectives were twofold. First, we aimed to generate a critical reflective systematic map describing the nature of qualitative approaches to voice UX. Second, we aimed to conduct a qualitative evidence synthesis of the qualitative findings from those studies about voice UX. Our research questions were threefold. For the first objective, we sought to map out the state of affairs in qualitative work on voice UX. We asked: *RQ1. What are the characteristics of the research and qualitative approaches used? (Methodology)* Our second objective was to understand why qualitative inquiry is used for voice UX and what characteristics of voice UX inspire, or potentially require, qualitative approaches. We asked: *RQ2. What reason(s) did the researchers provide for taking a qualitative approach to evaluating voice UX? (Epistemology)* For the third objective, we sought to demonstrate the value of using qualitative approaches to understand voice UX, aiming for a synthesis of findings, where possible, as well as highlighting those findings derived through qualitative inquiry. We asked: *RQ3. What findings on voice UX have these qualitative approaches generated? (Knowledge)* These questions generated a rich tapestry of methodological descriptions and knowledge synthesis on qualitative approaches to the study of voice UX, our main contribution. We hope that this work demonstrates the value of taking on a qualitative lens and will inspire further qualitative and mixed forms of inquiry within voice UX research.

## 2 METHODS

We conducted a systematic review based on the Cochrane standard [60]. Cochrane is a worldwide group of professionals who have developed standardized methods and procedures for conducting systematic review work that has the goal of comprehensively gathering and analyzing the available materials on a given subject or within a field of study [53]. While focused on the health and medical domains, the Cochrane standard is translatable to other areas [53] and often used in the field of human-computer interaction (HCI) [131], under which voice UX research and practice falls. The method requires the use of a standard procedure in review protocol and reporting. For this, we followed a modified version[2] of the Preferred Reporting Items for Systematic Reviews and Meta-Analyses (PRISMA) 2020 Statement[3], which is recommended by Cochrane [111]. The PRISMA statement is made up of a flow diagram representing the review process and a 27-item checklist that covers all action and reporting items considered foundational to the Cochrane systematic review method, e.g., eligibility criteria, search strategies, etc. As a tool, it supports customizations to enable use in other disciplines, like HCI. We start by defining terms and operationalizing key concepts. We then report on our protocol in detail. This protocol was registered before data collection at OSF[4] on June 11th, 2021.

## 2.1   DEFINITIONS AND FRAMEWORKS

We recognize the interdisciplinary nature of certain terms and concepts used in this work. For clarity, we define these terms and cite the conceptualizations we relied upon in our work.

### 2.1.1   *Voice and Speech.*

*Voice* is defined as "an expressive aural medium of communication" premised in sound [139]. Voice refers to the use of sounds to express *utterances*, comprising vocalizations, verbalizations, and other forms of auditory content (*output*), that are heard and interpreted by the sender and any recipients (*input*), including non-humans [47,79,80]. In humans, voice is typically generated through vibrations produced by the vocal folds found in the larynx that are then altered by features of the vocal tract, including the throat, nasal cavity, and so on. Voice can also be generated by machine-synthesized vibrations. While voice is often used to communicate words, i.e., the act of verbalizing, this is not always the case; babbling, coughing, whistling, backchanneling, vocal pauses and fillers, and other non- and pseudo-linguistic vocalizations are also matters of voice. *Speech*, while often used interchangeably with voice, is a distinct concept [104,116] that can refer to verbalizations as well as more technically to the state of voice after leaving the body. In this survey, we focus on work that explicitly targeted or assessed voice as the "how" of sound-based communication.

---

[2] Our deviations were due to the qualitative nature of the studies included and disciplinary differences between the medical and health sciences and computer science and engineering. For example, most publications do not use structured abstracts.

[3] http://www.prisma-statement.org

[4] https://osf.io/c6y8z

### 2.1.2 Vocalics and Paralanguage.

*Vocalics,* also called *paralanguage*, refers to the perceived *nonverbal paralinguistic* and *prosodic* qualities of voice [122]. Nonverbal *paralinguistic* qualities include tone of voice, loudness, pitch, and timbre (or voice quality). Nonverbal *prosodic* qualities include rhythm over the course of a vocalization, intonation as variation in pitch, and stress or emphasis of certain sounds over others. These properties vary and intersect to produce higher-level expressions indicative of emotion and affective states, social information (including personality, dialect, and gender), and biological states (including health and age) [80,85]. Given its importance to voice, we paid special attention to vocalics in the work surveyed.

### 2.1.3 Utterances and Conversation.

An *utterance* is an auditory message [112] or the "what" of voice-based communication [124]. Utterances can be *linguistic*, i.e., word- and sentence-based, or *non-linguistic*, i.e., gibberish, vocal bursts [14]. *Conversation* involves interactive rounds of utterances between at least two actors [25,112] but may or may not involve voice, as soundless, text-based CUIs and chatbots demonstrate. In this survey, we included technologies that used any kind of utterance but excluded technologies that did not have voice input and/or output.

**Artificial Intelligence (AI), Agents, Interfaces, Systems, and Spaces.**

Voice is being embedded in a large array of interactive systems and technologies. At present, we do not have sufficient evidence to meaningfully distinguish the UX that these systems offer in terms of voice [139]; indeed, one motivation of this survey is to gear research in that direction. We therefore included all options. We begin with definitions for these options to distinguish common underlying technologies and modes voice-based of interaction. *Artificial intelligence* (AI) is a broad and murky term, due to its prevalence across various technologies as well as ongoing debate about what "intelligence" should or can mean, for computers and in general [108]. A more useful framing may be *agent*, referring to any person or thing that can take action in some world, virtual or real. Agents are embodied in the world, acting on and within it, potentially with other agents or on its own, typically through some combination of sensors and actuators [100,114]. Yet, not all voice-based computer technologies can be classified as agents.

Agents are distinct from the user, implying a relinquishing of user control [147]. Interactions and control mechanisms can be found in objects, e.g., voice-based refrigerators, interfaces, e.g., voice-controlled vehicle operation, and spaces, e.g., smart homes. There are many "bodies," possible and uncharted [139]. We have taken a platform-neutral approach in this survey, including work that involves voice even when the computer is embedded and potentially unapparent for the user.

### 2.1.4 *Voice User Interfaces (VUIs), User Experience (UX), and Voice UX.*

*Voice user interfaces* (VUIs), also called *speech interfaces*, are computer-based technologies that use voice as the primary or sole modality during interaction with people [26,141]. Specifically, they are "*what* a person interacts with" [26:5, our emphasis], be it an interface, agent, display, environment, or a voice of unknown source and potentially "bodiless" [139]. Such systems are characterized by *voice interaction* or "verbal and auditory interaction with a computer-based system that uses voice-only input, voice-only output, and/or voice input-output" [141:2]. As such, VUIs typically employ speech recognition systems and synthesized speech to enable spoken language interactions through the auditory modality [26]. However, VUIs can be one-way displays, delivering content through words or other sound-based phenomena, and/or recipients of human speech or other inputs, auditory or otherwise. We must also recognize that many VUI systems are prototypes that involve a human operator, such as in Wizard of Oz systems where a human controls the TTS or even speaks on behalf of the VUI, while the user interacts unawares [24,63,120].

Interacting with VUIs and other voice systems gives rise to *user experience (UX)*. The International Organization for Standardization (ISO) defines UX as a "user's perceptions and responses that result from the use and/or anticipated use of a system, product or service" (ISO 9241) [66]. *Voice user experience* (voice UX) then refers to the forms of UX that arise during voice interaction with a voice system [141]. Notably, UX is subjective, an experiential construct that comprises affective responses and emotions, attitudes and beliefs, preferences and perceptions, and psychological and behavioural responses within and outside of the experience [66]. UX occurs regardless of the intentions or awareness of the designers, developers, researchers, and/or providers of the system. Whether evaluated or not, UX happens for the user in a given context featuring interactions with a system [66]. In general, UX is a complex phenomenon and factor of study [55,56,82,86] involving subjective and objective modes of inquiry, quantitative measures and qualitative descriptions, and patterns of attitudes and behaviours that may shift over time and differ widely among individuals. Previous work (e.g., [24,139,141]) has reviewed quantitative approaches to studying voice UX. The current review adds to this prior work by addressing the thus far understudied qualitative side. We only included papers that provided insights on the UX of voice systems, excluding, for example, technical reports and system test papers.

### 2.1.5  *Qualitative Research.*

Qualitative approaches to research focus on the *qualities* of experience. Where *quantitative* approaches involve, as a baseline, some form of measurement resulting in *numerical* data, qualitative approaches are concerned with collecting non-quantitative or non-quantifiable data that *describes*, often deeply, the nature of experience [29–32]. As Pluye and Hong [117] characterize it, the qualitative is about "the power of stories" while the quantitative is about "the power of numbers." Both forms of data are empirical in nature but are distinguished by the data collection tool and data types used. Specifically, non-numerical data in the form of words, sound, images, video, and so on, and non-quantifiable data representing qualities of an experience rather than their frequency and intensity. Furthermore, qualitative methods tend not to be concerned with making generalizable claims or prescribing a fixed view of reality, while quantitative methods often take such a positivist approach and aim for generalizations about people. Finally, qualitative research often takes place *in situ* so as to fully capture the experience in its natural context [35]. This is not a requirement of quantitative forms of research, which often take place in artificial settings, i.e., labs. These differences and the potential for each approach to enhance and inform the other have led to the uptake of *mixed methods research* [29,30]. Mixed methodologies, as the name implies, *mix* quantitative and qualitative methodologies and/or data collection and/or data in some way. The ways in and extent to which the mixture can happen varies widely. Nevertheless, it is common in HCI [161], as well as in multidisciplinary fields of study generally [29,31]. As such, we include both nominally qualitative *and* mixed methods research in this review, zoning in on the qualitative aspects in cases involving the latter.

Qualitative research can be broadly categorized in a *typology of five approaches*: narrative, phenomenology, grounded theory, ethnography, and case study [32]. We describe each below:

- In *narrative* research, the subject of study and/or method is in the form of a *story*, typically comprising *events* in *chronological* order and focused on *one* person's lived experiences [27]. For example, a narrative study on voice UX could follow the path of use by an older adult in the home over several seasons, capturing novelty effects, onboarding, errors and surprises, relationship-building and -breaking, trust development, and so on.

- In contrast, *phenomenology* focuses on generating a common understanding or "composite description" [32:45] of a specific phenomenon experienced by *multiple* people. Important is acknowledging and bracketing out the researcher/s' own understanding and experiences of the phenomenon and discussing philosophical assumptions about the fruits of the research, especially how each person's subjective experience may differ and how these experiences may or may not inform objective reality. A voice project attempting to create a gender-neutral voice, for instance, would consider how individuals perceive the genderedness of the voice or its lack, as well as how the researcher's own framing of "neutrality" may not match that of participants.

- *Grounded theory*, rather than being merely descriptive as with phenomenology, attempts to *generate a new theory*, or general explanation, of a process or (inter)action-based phenomenon that is "grounded" in the data collected from multiple people [155]. Typically, data collection and analysis occur synchronously and cyclically, with the researcher/s comparing and contrasting accounts of experience as the process plays out. For example, grounded theory may be used to generate a description of how varying the frequency or pitch of a voice that speaks gibberish is dependent on the underlying language upon which the gibberish is based, requiring several cycles through different languages with the same participants.

- In *ethnography*, the researcher/s centre their inquiry on a specific and identifiable cultural group or community whose members *directly share* an experience of the phenomenon rather than experiencing that phenomenon separately [9]. For example, a voice-enabled smart refrigerator could be given to people in an office space who may use it in different ways together and individually, at certain points in the day or week. Often, the researcher/s enter into the same spaces and even the daily lives of the group so as to capture shared patterns of behaviour and social organization in line with a predetermined *theory*. Continuing the example, a researcher might attempt to map out how the smart refrigerator fits into theories of planned behavior, observing how attitudes and social norms shift over time in response to the presence of the new "voice" in the room.

- Finally, *case studies* focus on a slice of experience: a "thick" description of a bounded entity, which could range from an individual to a community to a problem to a relationship and so on [174]. A case study of voice UX could, for instance, focus on the installation of a voice-based agent in place of a receptionist at a popular local hospital, and people's first reactions to it. Importantly, the case is strictly defined by specific parameters ("bounded") and followed over time at a particular site. A case is generally chosen as a representative means of better understanding and/or illustrating a complex, unique, and/or multifaceted phenomenon.

Importantly, these categories of qualitative inquiry are not exhaustive nor restrictive. HCI research that employs qualitative forms of inquiry can be liminal: involving a range of data collection, data forms, and data analysis that may not fit into typical research patterns. For example, Blandford [8] coined the term *semi-structured qualitative studies* to describe a common form of qualitative engagement in HCI that tends to use observation and interviews that are then systematically coded. Moreover, critical methodologists operating within and beyond HCI [8,28], notably Rode [129], have pointed out that qualitative approaches undertaken in HCI, even under the same name, may not be employed in the same way as in other disciplines. Finally, for these and other reasons, researchers may not label their approach explicitly, opting for a more general moniker, especially "qualitative study." As such, we use this five-part typology to describe the *qualitative orientations* of this body of work, focusing instead on the forms of data, information, and knowledge that may be usefully generated through such a variety of qualitative frames on voice UX.

Additionally, we do the same for mixed methods studies, harnessing Greene, Caracelli, and Graham's [50] five-part typology to describe the ways in which qualitative forms of inquiry were included. According to Greene, Caracelli, and Graham [50], methods may be mixed so as to *triangulate* for consensus, *complement* to enhance one or the other, *develop* the other form of inquiry in a follow-up study, *initiate* new questions or provocations that may buck against convention, and/or *expand* the scope of inquiry in a multi-part fashion. Where possible, we categorize the research as described by the researchers. However, in the absence of labels and to avoid prescription, we employ this five-part typology as a way of understanding *mixed methods orientations* to voice UX research. These framings thus represent a form of meta-level qualitative data that may be used to spark new forms of inquiry in future research.

## 2.2    ELIGIBILITY CRITERIA

Eligibility criteria were developed to identify voice UX research that involved qualitative modes of inquiry. Inclusion criteria were: Full studies, including user studies, experiments, interviews, field studies, etc.; qualitative approach/es used; voice UX as a main subject of study evaluated with a voice system (as defined above); and published to a peer-reviewed academic venue. Exclusion criteria were: Voice interaction not evaluated (even if the system has a voice or uses voice interaction); not enough detail reported to understand the study procedure; not studies, e.g., literature surveys, or not full studies, e.g., posters, proposals, small pilots; studies where only quantitative work was reported; inaccessible work (including after attempts to contact author/s directly); and work not in English, Japanese, German, or French, the languages known by the researchers.

## 2.3    INFORMATION SOURCES AND SEARCH STRATEGIES

General academic databases, specifically Web of Science, Scopus, EBSCOhost[5], and PsychInfo, and disciplinary academic databases, specifically ACM Digital Library and IEEE Xplore, were included in the search conducted on Thursday September 16th, 2021; see the spreadsheet on OSF[6] for the full search strategies per information source. We also manually added relevant papers from the data set provided by Seaborn et al. [139].

## 2.4    SELECTION AND DATA COLLECTION PROCESS AND DATA ITEMS

One researcher screened all abstracts, and a second researcher independently reviewed the papers marked for rejection. Then, four researchers independently reviewed the full text of about a quarter of these papers. Next, one researcher reviewed the papers flagged for rejection. Then, three researchers extracted the data, with the lead researcher taking on half of the papers and the rest divided equally. Extractions were checked by the lead researcher. If a paper was flagged for removal, all authors were notified beforehand. At each stage, the papers were presented in a random order. When disagreements occurred, the researchers discussed with each other and/or one of the researchers not involved until consensus was reached. Data items were: participant demographics, user groups, agent types and enabling technologies, algorithms, contexts of use, theories used, qualitative approach, qualitative data collection methods, instruments, inquiry methods, research purpose/s, reasons for using qualitative approaches, and qualitative findings.

---

[5] Note that we were unable to export records from EBSCOhost.

[6] https://osf.io/eb8n6

## 2.5 DATA ANALYSIS AND SYNTHESIS METHODS

We generated descriptive statistics, including mean (M), median (MD), interquartile ranges (IQR), standard deviations (SD), counts, and percentages, as well as inferential statistics (such as t-tests) for quantitative (such as gender data) or quantifiable (such as theme frequencies) data. Missing data was counted as its own variable. For the qualitative data, a reflexive thematic analysis approach was used [12] out of recognition of the diverse qualitative nature of our data set. Three researchers coded each set of data. The research purpose data was coded in a combined inductive-deductive way, inspired by the thematic structures from previous research on quantitative voice UX [24,139,141]. For example, we attempted to apply the dependent variable themes from Seaborn and Urakami [141], which included usability, engagement, cognition, affect, sociality, attitudes, perceptions of voice, and aggregate UX; however, not all of these categories applied to our data set. The themes that emerged were then used to structure the research findings data. A deductive approach was taken for the qualitative reasons data; one researcher used the frameworks provided by Creswell and Poth [32] for the qualitative studies and Greene, Caracelli, and Graham [50] for the mixed method studies, as outlined in Creswell and Plano Clark [31]. Finally, an inductive approach was taken for the instrument data, which was diverse and not tied to a theory or paradigm. In recognition of this, two researchers synchronously and collaboratively coded the data to develop a shared representation of its utility. For all analyses, codes were grouped into themes and theme frequency counts were generated. We did not conduct risk of bias assessments or assess certainty/confidence in reporting due to the qualitative and mixed methods nature of this body of work, as well as its variety, as assessment may be difficult or inappropriate, e.g., if the authors are philosophically opposed to measures of agreement [96].

## 3   RESULTS

From an initial 609 results across six databases, 66 items were selected for inclusion; refer to our PRISMA flow diagram, available on OSF[7]. Note that counts refer to papers unless otherwise specified. The full dataset is available on OSF[8].

### 3.1   METHODOLOGY: CHARACTERISTICS OF THE RESEARCH AND QUALITATIVE APPROACHES (RQ1)

#### *3.1.1   Participant Demographics and User Groups.*

Publications spanned 18 years, from 2003 to 2021, but most are clustered in the last five years (Figure 1). A total of 3080 people participated across all studies. The average number of participants per study was 46.7 (SD=108.1, MD=21, IQR=35.5). Breaking this down by qualitative inquiry method, we find an average of 36 (SD=34.3, MD=20, IQR=31) for interviews, 243 (SD=423.6, MD=40, IQR=231) for questionnaires (which could be large-scale surveys or questionnaires deployed in a study, hence the large range), 44 (SD=38.6, M=38, IQR=14) for focus groups, 20 (SD=11.3, MD=20, IQR=8) for workshops, 32 (SD=23.5, MD=21, IQR=33.3) for observational studies, and 20 (SD=9.2, MD=19, IQR=3) for transcript analyses; note that counts overlap because one study may have used multiple modes of inquiry. 1251 women (42.4%) and 1317 men (44.6%) as well as three with no gender identified were reported; no one of another gender identity was reported. A t-test did not find a statistically significant difference in the number of men and women, indicating balanced representation among men and women, $t(2567) = .063$, $p = .47$. However, the gender of 512 participants was not reported, so this result is tentative. Most participants were drawn from the general population (24 or 36.4%) or constituted a special group (21 or 31.8%). Eleven studies (16.7%) included older adults, ten (15.2%) included children, and nine (13.6%) included students. Two studies did not specify the user group. Six studies focused on caregivers and/or mothers, four on blind/low-vision/visually impaired folks, three on those with or were parents of those with HIV, three on employees, two on roboticists or designers, and two on people with dementia.

---

[7] https://osf.io/3arz9

[8] https://osf.io/yhwrs

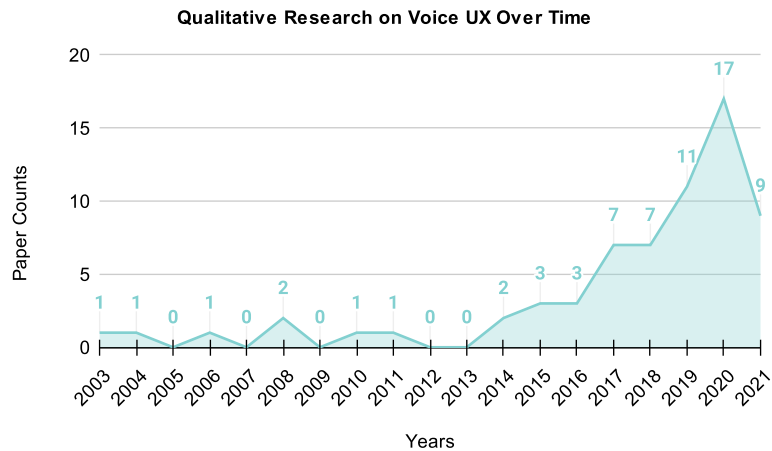

*Figure 1 Papers employing qualitative inquiry to explore voice UX over time.*

### 3.1.2 *Voice Systems and Enabling Technologies.*

Note that many studies used more than one system, so count totals may vary. Most voice systems (Figure 2, left) were smartphone-based (19 or 28.8%), general phones (17 or 25.8%), and robots (15 or 22.7%). Other voice systems included virtual characters (6 or 9.1%), computers, i.e., computer voice through speakers (5 or 7.6%), or some other computer-based system, such as custom prototypes (7 or 10.6%). Notably, four studies (6.1%) did not report on a type of voice system, due to the nature of the studies as attitudinal, explorative, and/or participatory. Specific technologies included interactive voice response systems (16 or 24.2%), smart things, especially speakers (13 or 19.7%) but also cars (2 or 3%) and watches (2 or 3%), and the NAO robot (2 or 3%). TTS's included Amazon Alexa (12 or 18.2%), Google Home (6 or 9.1%), and NAO (3 or 4.5%). Voice-enabling technologies included TTS's and recordings (Figure 2, right). Most studies did not report the TTS used, if any (16 or 24.2%) while fourteen (21.2%) used another TTS, i.e., one-offs, and five (7.6%) used a custom TTS. Five studies (7.6%) used pre-recorded speech and one used non-speech sounds. Overall, one-third (22) studies looked at human voices, one-third (22) at synthetic voices, and one-third (22) did not report on the voice. Two studies used animal voices and two did not specify the voice type. While 29 studies (43.9%) did not specify the underlying intelligence, six (9.1%) used Google, notably DialogFlow, nine (13.6%) used Amazon Web Services (AWS), eight (12.1%) used a custom system, and six (9.1%) used another platform, such as NaoQi.

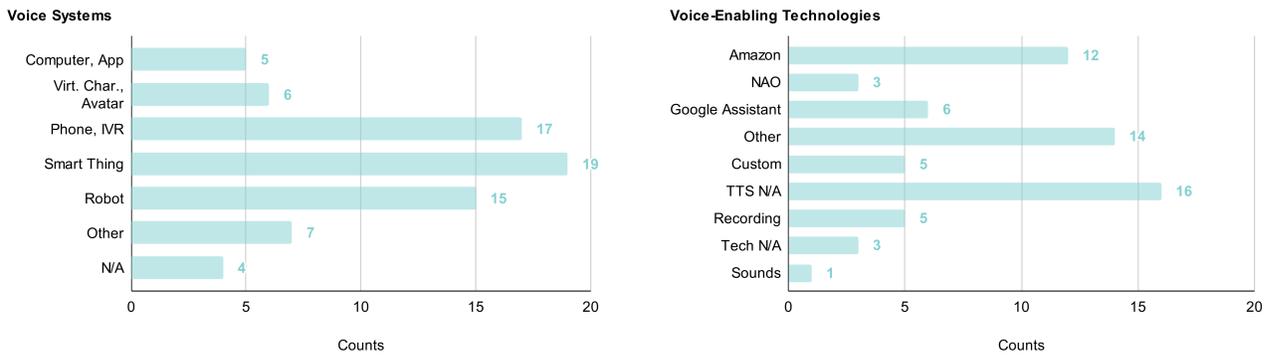

*Figure 2. Voice systems (left) and voice-enabling technologies (right).*

### 3.1.3 Contexts of Use.

Most studies explored an instruction context (24 or 36.4%), conversation (15 or 22.7%), or a service context (14 or 21.2%). Others employed a game (6 or 9.1%), a driving context (4 or 6.1%), or another context (8 or 12.1%), such as controlling a smart lock, disclosures of social behaviour, and folding a cloth.

### 3.1.4 Purpose for the Research.

We derived four high-level themes that describe the reasons behind conducting qualitative research on voice UX (refer to Table 1 for a summary; the full table is available on OSF[9]). Purposes expressed by the researchers were experiential (63 or 95.5%), targeting usability (30 or 45.5%) and/or embodiment (33 or 50%), attitudinal (45 or 68.2%), communal (41 or 62.1%), including specific populations (30 or 45.5%) and social factors (11 or 16.7%), and vocal, or reasons deemed of particular relevance, if not exclusive to voice UX (17 or 25.8%). This last category, which may be especially pertinent here, included: conversational styles [128], response styles [177], verbal behaviour [125], communicative signals [136], voice input/output (voice input/output) [166], disclosures [93], long-turns in conversation [33], living noise [70], source orientation [51], conversation expectations [25], emotion in conversation [83], features of dialogue [121], verbal prompts [1], synthetic versus real voices [81], conversational exchange and repair [39], and "rich" communication [3].

---

[9] https://osf.io/qup5j

*Table 1. Thematic framework of research purposes.*

| Theme | Sub-Theme | Description | Examples | Citations |
|---|---|---|---|---|
| Experiential | Usability | Relates to ease of use, effectiveness, and efficiency, including accessibility, alternative modalities, performance, and obstacles. | "explored general UX" … "how older adults use the system" … "effect on adherence behaviour" … | [1,11,13,18,21,33,39,42,52,61,73,74,98,99,103,123,128,130,156,159,163,168,169,173] |
|  | Embodiment | Relates to how the agent was or could be embodied in form factor and interaction, including (multi)modality, humanlikeness, and nonverbals. | "explored 3 types of displays from voice to visual" … "expressiveness in speech patterns" … "sought to identify gaze behaviours" | [1,3,6,25,34,37,54,61,74,81,83,91,113,119,121,125,136,143,144,146,154,156,171,173,176,177] |
| Attitudinal | n/a | Relates to attitudes and ideas expressed about the system, including acceptability, mental models, and trust. | "understand feelings of trust" … "how appropriate kids thought the agent was" … "understand perceived persona" | [1,5,6,10,13,25,34,37,40,42,51,52,54,57,64,70,73,81,88,90,91,93,99,113,115,121,123,125,126,130,134,136,144,154,159,162,163,168,171] |
| Social | n/a | Purposes were specialized by population and social factors. | "rural mothers" … "looked at family use" … " | [5,6,10,13,18,37–40,42,52,54,57,61,64,70,72,73,93,98,99,115,121,123,126,130,134,143–145,154,156,159,162,163,166,172,176] |
| Vocal | n/a | Relates to factors particular or exclusive to voice UX. | "hands free design" … "explore long-turns in conversation" | [1,3,7,25,33,39,51,70,81,83,93,121,125,128,136,166,177] |

## 3.2 THEORETICAL FRAMEWORKS AND CONCEPTUAL MODELS.

Many studies (27 or 40.9%) did not report on theory. Six (9.1%) used the Technology Acceptance Model (TAM) [167], four (6.1%) referred to the Computers Are Social Actors (CASA) paradigm [87,106,107], and two referenced the Uncanny Valley model [101]. Others were found once: the Big 5 theory of personality [71], the circumplex arousal-valence model of emotion [133], interpersonal communication theory [77], theory of mind, the integrative model of organizational trust [94], personification and the three-factor model of anthropomorphism [43], rapport [158], hyperarticulation [152], source orientation [51], peer instruction [95], the circumplex arousal-valence model [133], interpersonal communication theory [132], trouble-source repair [110], theory of common ground [23], media equation theory [105], cognitive load theory [157], consequential communication [142], para-social relationships theory [62], social response theory [105], extended privacy calculus model [36], and the stereotype content model [46].

### 3.2.1 Inquiry Methods and Modes of Qualitative Inquiry.

Most studies (39 or 59.1%) relied on a qualitative approach. A further 21 (31.8%) used a mixed methods approach, while six (9.1%) generated qualitative data using quantitative methods. The most frequent mode of inquiry was interviews (43 or 65.2%). Other common modes included observation (26 or 39.4%) and field studies or "in the wild" research (27 or 40.9%). Others included focus groups (7 or 10.6%), questionnaires and surveys (7 or 10.6%), transcript analysis (5 or 7.6%) and design workshops (2 or 3%). Techniques used once included diaries [70] and drawing [88]. Most studies evaluated participants' direct interaction with a voice system (34 or 51.5%). Many studies were conducted in-field rather than in the lab (24 or 36.4%) while others used Wizard of Oz regardless of the study context (13 or 19.7%). Ten (15.2%) of studies used a scenario approach, with six (9.1%) using video and four (6.1%) using audio. Six (9.1%) did not describe in detail. A further five (7.6%) used another method, such as conducting a simulation, analyzing drawing, and sourcing transcripts.

### 3.2.2 Qualitative Instruments.

While a range of inquiry methods and modes of inquiry were used, many researchers employed questions to gather qualitative data. We found this to be the case in 46 (69.7%) papers. We then inductively identified 331 codes across four themes and eight sub-themes bearing on how and why qualitative instruments were used. Table 2 represents the distribution of these alongside the overarching qualitative approaches deployed and the citations touching on each of the categories identified; the full data set with tool extractions is available on OSF.[10] Usage refers to current or previous usage (12 codes or 3.6%). Phenomenal impressions refer to interactive experiences (95 codes or 28.7%), agent impressions (42 codes or 12%.7), and attitudes towards the voice system (59 codes or 17.8%). Methodology includes feedback elicitation, co-design strategies, tool validation, data mixing, and type of respondent, such as the user, a co-designer, or a system-manager (96 codes or 29%). We finally considered lack of rigor in reporting, via ambiguous or missing details (26 codes or 8.2%). For example, Strathmann, Szczuka, and Krämer [154] wrote of asking "63 questions" on "the transparency of the voice assistant's functionality, the intensity of the relationship built, as well as the communication behavior towards the device" (p. 6). While they provided an example question for each category, e.g., "*Would you tell Google/Alexa a secret?*," they did not specify an instrument or otherwise provide all questions used. These gaps and diversity across the corpus of papers makes it difficult to generalize best practices. The paucity of rigorous studies highlights the opportunities for development of qualitative tools for voice UX.

---

[10] https://osf.io/dj6cb

*Table 2. Thematic framework of qualitative research instruments.*

| Theme | Sub-Theme | Description | Examples | Citations | Qualitative and Mixed Methods Approaches |
|---|---|---|---|---|---|
| Usage | n/a | Previous or current usage patterns. | "we asked about familiarity with voice systems" | [6,7,13,34,52,57,113,136,156,166,169] | Phenomenology, Grounded Theory, Ethnography, Complementary |
| Phenomenal Impressions | Interactive Experience | General usability and UX as well as for the agent, obstacles and barriers, context of use, and social context. | "overall impression of the agent's facilitation" | [1,3,6,7,10,13,21,25,34,37,42,51,52,54,57,61,70,72,75,81,90,99,113,115,121,123,128,130,136,145,146,154,156,159,162,163,166,168,169,171] | Phenomenology, Grounded Theory, Ethnography, Case Study, Triangulation, Complementary, Expansion |
|  | Agent Impressions | Communicative ability and competency, CASA factors (or "identity" factors), and body factors. | "impressions of the agent" … "are they talking to a phone or assistant" | [3,5,7,10,13,21,34,51,57,72,72–75,81,90,99,113,115,121,123,126,128,136,146,154,156,162,163,168,171] | Phenomenology, Grounded Theory, Ethnography, Triangulation, Complementary, Expansion |
|  | Attitudes | Utility and usefulness, acceptability and willingness-to-use, preference, and trust and security. | "is the VA useful" … "do you have any privacy concerns while using it" | [5,6,10,13,21,25,34,37,52,54,57,61,70,72–74,81,90,99,115,121,123,126,130,136,145,146,156,159,163,166,168,169,171] | Phenomenology, Ethnography, Grounded Theory, Complementary, Expansion |
| Methodology | Feedback Elicitation | Information-gathering that does not raise to the level of co-design. | "suggestions to improve" | [3,6,10,13,21,25,34,61,70,73,74,81,90,98,99,113,115,121,145,159,162,163,168,169] | Phenomenology, Ethnography, Grounded Theory, Complementary, Expansion |
|  | Co-Design | Users involved as co-designers. | "what features to add" | [3,10,52,61,72,73,75,98,99,145,146,156,169] | Phenomenology, Ethnography, Complementary |

| Theme | Sub-Theme | Description | Examples | Citations | Qualitative and Mixed Methods Approaches |
|---|---|---|---|---|---|
| | Tool Validation | Focus is on validating a research tool. | "how did you use the tools to explain your view" | [75,90,136] | Phenomenology |
| | Data Mixing | Varieties of qualitative and/or quantitative data collected together. | n/a | [5,37,42,72,73,90,99,126,130,166] | Ethnography, Case Study, Triangulation, Complementary |
| | Type of Respondent | User, user-adjacent stakeholder, experts, providers. | "We then interviewed parents of child users" | n/a | Phenomenology, Ethnography, Grounded Theory, Case Study, Complementary, Expansion |
| Ambiguous Reporting | n/a | Details about the instrument missing, such that they cannot be extracted or reproduced. | n/a | [1,3,6,7,13,21,25,37,42,51,52,54,61,70,72,75,98,113,126,130,146,154,156,166,168,171,177] | Phenomenology, Grounded Theory, Ethnography, Case Study, Triangulation, Complementary |

### Epistemology: Reasons for a Qualitative Approach to Researching Voice UX (RQ2)

Researchers provided a range of reasons behind employing a qualitative approach and the specific qualitative approach taken. Table 3 presents the methodological orientations based on our thematic analysis using the typologies from Creswell and Poth [32] and Greene, Caracelli, and Graham [50]. Note that while about one-quarter (17 or 25.8%) of researchers provided explicit reasons and orientations to qualitative inquiry, a greater portion (30 or 45.5%) were less exact, such as by not describing with enough detail or not referring to specific qualitative typology. A further 19 (28.8%) did not provide any reason or explanation of their orientation to inquiry, simply stating that a qualitative approach was used or merely reporting on the use of a qualitative means of inquiry, such as an interview.

Of the qualitative-only work, none used a narrative orientation. Most took a phenomenological (21 or 31.8%) or ethnographic approach (17 or 25.8%), five used grounded theory (7.6%), and one [42] was a case study. Of the mixed methods research, most used qualitative inquiry in a complementary way (15 or 22.7%) or for triangulation with the quantitative data (8 or 12.1%), while others sought to expand the range of inquiry (4 or 6.1%) or had another reason for mixing methods (2 or 3%), specifically as a corrective response to failures in quantitative procedures [18] and based on prior recommendations in the literature [5]. None had developmental or initiation-oriented reasons for mixing methods.

The reasons provided were diverse and hard to categorize with more specific themes. Moreover, the level of depth and detail varied widely. Bailey et al. [5] followed a recommendation by other researchers to use qualitative methods. Ellway et al. [42] went into great detail, framing their approach as one informed by interpretivism in information system fields and a desire to answer "how" questions. They further explained that they chose one, rather than multiple, cases for three reasons relating to complexity of the experience, the multiple actors and roles involved in the call centre, and the ability of the method to allow the researchers to focus on how calls were passed between teams, a key factor in their research. Damen et al. [34] and Shamekhi et al. [146] explained their use of a complementary mixed methods approach as a means of generating greater insight into the quantitative data analyses. Bentley et al. [7], working with a large base of 1000 commands, sought to identify patterns within and among teams and team members, effectively combining phenomenology and grounded theory. Lopatovska and Williams [91] aimed to understand how people personified Amazon Alexa, a qualitative data form that may have similar features across people, especially with the focus on one agent. Similarly, Lee, Kim, and Lee [88] conducted a drawing study on several agents but also several different people. Robb et al. [128] noted that they could focus on topics of interest while also shifting focus to emergent participant perspectives. Sayago [134] recognized that being a participant and observer in the classroom would allow for more naturalistic observations. Sezgin et al. [145] wished to better understand the quantitative data previously collected and how it might be used in future systems. Reasons were diverse but practical, focusing on the merits of the methods.

*Table 3. Studies categorized by orientation to qualitative inquiry.*

| Approach | Typology | Counts (%) | Citations |
|---|---|---|---|
| Qualitative | Narrative | n/a | n/a |
| | Phenomenology | 21 (31.8%) | [1,3,7,21,25,38,39,51,61,75,83,88,91,103,121,136,143,154,169,172,176] |
| | Grounded Theory | 5 (7.6%) | [38,51,70,103,128] |
| | Ethnography | 17 (25.8%) | [6,10,13,54,70,72,73,98,123,128,130,134,144,159,162,163,168] |
| | Case Study | One | [42] |

| Approach | Typology | Counts (%) | Citations |
|---|---|---|---|
| Mixed Methods | Triangulation | 8 (12.1%) | [33,37,40,64,99,113,171,177] |
| | Complementary | 15 (22.7%) | [5,11,34,52,57,74,81,90,93,119,126,145,146,156,166] |
| | Development | n/a | n/a |
| | Initiation | n/a | n/a |
| | Expansion | 4 (6.1%) | [74,115,125,173] |
| | Other | 2 (3%) | [5,18] |

## 3.3 KNOWLEDGE: QUALITATIVE FINDINGS ON VOICE UX (RQ3)

We now turn to what knowledge has been garnered through qualitative approaches to evaluating voice UX. We organize the findings by *domain of study* to highlight the state of knowledge for specific areas of study, application domains, or topics of focus. Each subsection represents a domain of study: blind and low-vision users and accessibility; teaching and instructional contexts; driving and navigation; well-being and therapeutic contexts; health information contexts; multi-user contexts; user mental models of anthropomorphism and social identity; user mental models of embodiment and form factors; and vocal factors. Although diverse, within each domain of study we can summarize the findings by way of a series of *voice UX factors*, described in the following high-level framework:

- *Experiential factors*, which include:
    - *Usability* refers to how the agent, interface, space, or system performed for the user at the task or activity, including accessibility and obstacles. Usability, as an ISO standard, is a basic part of evaluating UX in general [67] and refers to meeting user needs and desires for effectiveness, efficiency, and satisfaction.
    - *Agent embodiment* refers to how the agent is situated within, interacts with, and is influenced by the context of use, including the environment and other agents, human or otherwise, within it [100]. Embodiment typically refers to the "body" of the voice UX system, including sensory modalities, enabling technologies and devices, form factors and morphologies, notably level of anthropomorphism or humanlikeness, and expressiveness. When it comes to voice UX,

vocalics may occur alongside visual and tangible modes of expression, including nonverbal behaviours and gestures [139,165].

- *Attitudinal factors*, which include:
    - *Acceptability* refers to the attitudes that people have about the voice system, from adoption to preferences, liking and disliking, feelings of satisfaction and appropriateness, and general usefulness for the system in their lives. Acceptability may be especially important for voice as a new and evolving mode of interaction and system expression for most people [167].
    - *Trust* and *privacy* are entwined attitudinal factors that have a long history in agent-based and general system work involving computation [68,87,92,164]. Trust refers to opening up to others with the expectation of a reliable and honest exchange, while privacy refers to closing off from others or restricting access in some way. Subfactors include rapport, safety, security, reliance, abuse, betrayals and being tricked, and so on.
- *Social factors* or *sociality*, which refer to the social features and relations involved in or implicated by voice UX. Voice is essentially about communication, a social act likely to have social repercussions, especially in terms of creating feelings of social connectedness among people.
- *Vocal factors* that refer to voice-specific and potentially voice-exclusive forms of interactions and effects.

We further contextualize this framework with respect to specific qualitative modes of inquiry in Table 4. For now, we consider what we know (and perhaps do not yet know) by domain of voice UX research. This is not an exhaustive list of domains; rather, it represents the largest clusters of related work in the corpus of work surveyed.

### 3.3.1 Blind and Low-Vision Users and Accessibility.

Voice may be particularly relevant to blind/low vision users. Voice-only modes of interaction were found to be beneficial and enabling, even when compared to non-visual alternatives, such as tangible interaction [18,98]. Metatla et al. [98] conducted a series of co-design workshops with students, educators, support staff, and local authorities, a novel approach to identifying stakeholder-sensitive insights on designing voice systems for school contexts. Vashistha et al. [166] identified voice interaction as an essential mode of accessing and expressing oneself on social media. Visually impaired older adults appreciated the notion of a medical voice agent, even though the implementation was not as smooth or humanlike as desired [54]. Indeed, they had mixed reactions to the medical voice agent [54]. Speed and pronunciation of medication names were key issues, as a desire for a smaller device as well as direct interaction with people rather than technology. Most of these issues may be resolved with advances in the underlying technologies. In sum, voice UX offers a compelling and accessible experience for people with impairments related vision, even if it is not perfect yet.

### 3.3.2  *Teaching and Instructional Contexts.*

Voice systems have been deployed as teaching aids across a variety of contexts, from helping older adults how to use an electronic programme guide [18] to learning how to use a text editor [171] to simply being a common mode of engagement with smart speakers, i.e., information seeking [7,134,144]. Others have focused on the ability of the technology to enable an educational experience compared to human standards in terms of task performance [40,136], notably as "peers" or "companions" [40,134] or even "supporters" or "sponsors" in the context of behaviour change [42,73,126]. In learning contexts, trust has been a focal point, notably achieved by way of vulnerability through disclosures [93] and characters speaking in the local language [64,72]. In general, voice-based engagement was preferred over other options, with notable caveats. Device trade-offs may be important to consider for new learners. Militello et al. [99], for instance, found that phones were easy to learn but prone to speech recognition errors, while the more accurate smart speakers were too expensive. Multimodality might also be important for learners and learning contexts, with many participants expressing a desire for visuals or even tactility to add voice-based engagement [98,99,145,173], perhaps especially when the voice is coming from a robot [40]. Children may be a special category of student in need of special care because of the linguistic engagement offered by many voice-based systems. Du et al. [39] found that Alexa was too erudite during learning games and parents had to intervene. In contrast, Dönmez et al. [38] found that children bonded to a robotic tutor that was playful, provided clear responses and constant feedback, although it is unclear if this was eased by the robotic embodiment of the agent. On the other hand, Bailey et al. [5] found that low-literacy adult learners preferred the interactive voice response system to events involving other people, perhaps because of social stigma around literacy.

In sum, voice appears to be an acceptable and sometimes powerful mode of engagement in instructional settings with a variety of learners. Still, we should consider the inclusion of visual or other modalities, choose the enabling device and persona based on the socioeconomic background of the learners, and gear the linguistic ability of the voice to the learner, especially when they are children, low literacy, or not fluent in the system language.

### 3.3.3 *Driving and Navigation.*

Voice interaction within vehicles and navigation contexts is an emerging area of study that can involve lab and field work, or some combination of the two, for experimental control or safety's sake. The surveyed work raised certain factors as particularly relevant, if not crucial, for these contexts. The clear benefit of voice interaction in these contexts is the hands-free potential of voice interfaces. What form voice interactions can and should take have been explored and illuminated. Perrin et al. [113] found that understandability in the case of navigational information was best for voice feedback compared to visual, auditory, and tactile feedback modalities, even though auditory feedback reaction time was longest. Large et al. [83] found that input commands were performed in a command-based style. Others have considered the psychological and cognitive dimensions of voice UX. Kim et al. [74] found that voice-task demand was related to driver interruptibility and adaptive behavior such as reducing speed in a driving task. This may be attributed to the increased cognitive load caused by voice-only interaction. Braun et al. [11] found that the affective qualities of the voice assistant were beneficial for expressing empathy with the driver and mediating their mood, especially negative states. In sum, the qualitative findings on voice UX in driving and navigation contexts highlight the usability of the auditory modality and its flexibility for the task, as well as provide support for voice against driver distraction and road rage. Voice in driving and navigation contexts may be best designed around short commands that do not induce additional cognitive load or distractions on the driver and personas that express empathy in a way that adapts to driver mood.

### 3.3.4 *Well-being and Therapeutic Contexts.*

Voice UX in well-being and therapeutic contexts is an emerging domain of study. Voice-based interaction in therapeutic contexts can improve mood and relaxation for pain relief [123]. Emotional voice assistants that empathize with the user improved negative emotional states and were rated positively [11]. Voice was also perceived as having a calming effect and acted as social support that helped with smoking cessation [126]. Clearly, voice interaction can supplement therapeutic activities. Other work has brought attention to factors of sociality and anthropomorphism of the voice system. In Huang and Li [64], "role play" with characters in the local Cambodian language built trust. Pou-Prom et al. [121] found that human voices were more engaging and that the Wizard of Oz prototype, being driven by a human operator, allowed for greater flexibility in responses, which was highlighted as particularly well-received by participants compared to the autonomous, inflexible version. Other research highlights opportunities for interaction design work. For instance, participants in a study on therapy adherence monitoring desired to edit inputs with voice [52]. Altogether, voice can be a calming and empathic supplementary feature, but must otherwise be designed for the particular well-being activity or therapeutic context, with the needs of the care recipient in mind.

### 3.3.5 Health Information Contexts.

Most work has been focused on health information purveyance, especially by way of interactive voice response systems in low-literary and low-resources areas of the world. Key factors include expectations of system performance, especially length of call (shorter is better), authenticity, repetition, effort and engagement, social influence and sharing of devices, and fear of making mistakes or unintentional disclosure [5,13,42,52,130]. Voice was a better fit than text for low literacy folks [64] and blind/low-vision folks [156]. Interactive voice response systems were also preferred over SMS in mHealth applications [130]. Still, Militelo et al. [99] found that a visual display improved satisfaction with the voice system. Social identity may play a role, as using the voice of a real doctor led to increased perceptions of usefulness and trust for people living with HIV [72]. Social context may also be crucial for usability in community [52] and familial [145] contexts; at the least, we must recognize that people may share their voice system with others or use it together on purpose. Overall, the voice modality was superior to other options from usability and acceptance perspectives, but, given the possible sensitivity of health information, caution must be paid when multiple users (or bystanders) may be involved.

### 3.3.6 Multi-User Contexts.

Privacy concerns [34,40,130,159] are paramount in multi-user and public contexts. Trajkova and Martin-Hammond [159] and Adaimi et al. [1] reported hesitation to use VUIs in the presence of others or public, shared spaces. The social robot Fribo was tested in a four-week field study, which showed that frequent real-world social interactions through activity sharing lessened the feeling of privacy intrusion [70]. A voice system sharing "living noise" and daily activity information with close friends improved feelings of social connectedness between them [70]. Using multiple voices to generate the impression of multiple agents increased the sense of being in a team, even though a single agent was quicker and easier to remember [176]. Age must also be considered; for example, appropriate speech patterns in a voice system for children were found to position the VA as a near-peer [6]. Relatedly, Haberer et al. [52] found that parents of HIV-positive children shared their phones with each other and potentially other people, indicating that we should be careful about relying on a "solo user" model even with "personal devices." The ways in which voice UX plays out in multi-user and multi-agent contexts may be a key trajectory in future research. In short, voice systems may be used socially or affected by the social context, with longer-term engagements easing users away from privacy concerns—for good or ill.

### 3.3.7  *User Mental Models of Anthropomorphism and Social Identity.*

Voice appears to be crucial for establishing an agent's social identity and humanlikeness. Even when we know better, we often cannot help but attribute social and human-like characteristics to the machine [115]. People seem to prefer natural, human-like voices over artificial agents [81,121]. The use of natural speech can lead to anthropomorphizing [154]. With Alexa, users tended to use polite expressions such as "thank you" and "please," similar to natural conversations [91]. Kim et al. [75] identified personality traits that people desire in voice agents: empathizing, trustworthiness, submissiveness, and smart-yet-modest, as well as distinctive traits that express individuality and neutral traits, such as adapting to the preferences of the user. Pitardi and Marriott [115] showed that perceptions of privacy and trust were formed based on the brand and producer behind the VA (e.g., Alexa being made by Amazon). Perceived privacy may thus depend not only on the design of the VA but also on the brand image connected to it. Personalization of the voice interaction might increase acceptability. For example, Doering et al. [37] reported that users wanted to be recognized by the system, and the personalized version was preferred in a study by Kassavou and Sutton [73]. However, individual factors seem to play a role; for example, a system with a personality was perceived well by some but not by others [90]. Altogether, people have a tendency to anthropomorphize, especially when a human feature like voice is present, and we must take care to understand what mental models of agency and social identity this invokes—and then whether or not to modify properties of the voice to shift those models for the user, task, and/or context.

### 3.3.8  User Mental Models of Embodiment and Form Factors.

The physicality of the system in terms of embodiment and form factor seems to be important for optimal UX. However, whether physicality is advantageous or not seems to depend on its quality and the purpose of the interaction. Beneteau et al. [6] noted that close physical and social proximity helped to instill trust in the system. However, Pollman et al. [119] found that a robot's body motions were perceived as being confusing and concluded that these should be avoided if they cannot be improved. Similarly, Shamekhi et al. [146] found that the value of having a VUI with a face depended on the type of assistance that the system provided. Doering et al. [37] noted that users found it easier to accept the voice over the body of their humanoid mobile shopping robot. Furthermore, Damen and Toh [34] studied the role of voice gender, concluding that it was more important "what" something was said than "how" (e.g., gendered voice). Guzman [51] explored what the "source" of the voice was perceived to be, finding that most identified the source as the VA itself or the enabling device. Relatedly, participants in Pitardi and Marriott [115] were aware that Amazon, rather than the agent per se, was "listening in" during interactions with Alexa. Lee et al. [88] found that participants often included the speaker form factor within the agent design when drawing the "body" of voice-based CAs.

The voice system's physical attributes might also be key to the perception of robots as socially present agents [40]. Edwards et al. [40] reported that observable nonverbal codes were important to convey affect alongside voice. Seo et al. [143] identified a range of nonverbal cues accompanying voice during rapport-building and rapport-breaking scenarios with an industrial robot. Voice interaction supported verbal and non-verbal rapport-building, resulting in compliments but also in terse speech and even silence, indicating that the VA was treated like a social agent [143]. Ho et al. [61] showed that the combination of movements and verbal commands of a robot increased its social presence. Additionally, grade of assistance by way of voice seems to be another factor [171]. Physicality may also offer communicative modalities through nonverbal cues [165]. Nevertheless, this may not be universal or accessible for blind/low-vision user groups.

In sum, the presence or absence of a "body" for the voice and that body's form factor can have ramifications for trust, understandability, applicability to the task and context, and acceptance. A variety of "bodies" are available or imaginable against the range of task and contexts applicable to voice interaction. Determining ideal—and not-so-ideal—intersections is an undertaking that requires a mix of qualitative and quantitative approaches.

### 3.3.9 Vocal Factors.

One of our aims was to determine whether and how voice UX is distinguished from UX in general. Many of these are related or adjacent to conversation and speech, but not always. For example, Schreitter and Krenn [136] identified characteristics of dialogues by examining how people verbally teach a robot, including backchanneling, like "ah," sentence fragments, repair statements, contractions, repetitions, interruptions, the use of time markers like "now," and verbal references to the environment. In the overall corpus of work surveyed, highlights relevant to voice include:

- backchanneling [136]
- prosody, including rhythm and monotony [57]
- humanlike versus synthetic voice qualities [81]
- vocal expressions of im/politeness [125]

Speech-level features and utterances (speech content) appear to intersect with voice in important ways, as well. For instance, Martelaro et al. [93] found that a combination of vulnerable speech content and paralinguistics led to a greater number of personal disclosures from participants. Highlights from the surveyed work relevant to speech include:

- dialogue-level features [121,136,177]
- time, including length of exchanges [162], long-turns [33], and time of day [7]
- disclosures [93]
- reassurance [128]

Embodiment remains a key issue for voice systems, with several factors of embodiment unearthed or perhaps best explored with qualitative methods. Key factors in embodiment include:

- source orientation [51]
- richness of communication through supplementary modalities [3]

Qualitative voice UX work also revealed several topics relevant to interaction styles. Highlights include:

- hands-free interaction [99,128]
- multi-tasking via other modalities [1,99,128,144]
- hands-free interaction [1] enabling multitasking [99] (intersection)
- voice enabled by voice [166]

Communication maintenance also arose as a key topic. We hazard to characterize this topic as "relational," given that systems do not yet have the capacity for partner modelling, empathy, emotion, or other hallmarks of independent social intelligence. Still, voice UX often involves conversational formats and activities, and maintaining communication in the short and long terms, especially when problems arise or when users misbehave, is of ongoing concern. For example, Rehm [125] found that, when placed in a situation where the voice-based agent was deceptive, some participants responded with verbal abuse and other impolite, if not unacceptable (at least among people) behaviours. We highlight:

- habit- and relationship-building [40] and future expectations for the relationship with the system [25,83]
- negative and possibly abusive reactions [125]
- conversation exchange and repair [39]
- missed, unintentional, and unfamiliar intents and prompts [103,169]
- other issues with voice-based system performance (e.g., errors, missed activation) [40,42,103,169]

This likely does not represent the full range of voice UX factors that qualitative research can reveal. But it shows how qualitative inquiry can unearth and richly describe factors that may otherwise be missed.

## 4  DISCUSSION

Qualitative approaches are rapidly being taken up, as indicated by the spread of the work over time and the recent upward swing (Figure 1). This echoes the "vocaloid shift" recently identified for quantitative work on voice UX [139]. What does the qualitative side add? In Table 4, we summarize the relationship between factors of voice UX and the qualitative approaches used in the corpus of work so far. We now turn to making sense of the state of affairs, focusing on the patterns of experience and notable phenomena highlighted by this body of work. We discuss the value of qualitative approaches for voice UX in light of researchers' reasons, orientations to qualitative inquiry, the techniques employed, and the outcomes so far mapped out. We finally turn to the weaknesses, gaps, and opportunities identified through this review and how to address them in future qualitative work on voice UX.

*Table 4. Summary of relationship between voice UX factors and qualitative approaches.*

| Voice UX Factors | Experiential | | Attitudinal | | Social | Vocal |
| --- | --- | --- | --- | --- | --- | --- |
| | Usability | Agent Embodiment | Acceptability | Privacy & Trust | | |
| **Qualitative** | | | | | | |
| Narrative | | | | | | |
| Phenomenology | [1,21,39,61,103,169] | [1,3,25,61,70,75,83,88,91,121,136,143,154,176] | [1,70,154] | | [38,39,61,70,121,143,154,172,176] | [1,3,7,25,51,83,121] |
| Grounded Theory | [103,128] | [70] | [70] | [70] | [70] | [7,51,128] |
| Ethnography | [13,73,98,123,128,130,159,163,168] | [6,54,144,159] | [6,13,54,73,130,144,162,163,168] | [70] | [6,10,13,54,72,73,98,123,130,134,144,159,162,163] | [128] |
| Case Study | [42] | | [42] | | [42] | |
| **Mixed Methods** | | | | | | |
| Triangulation | [33,99] | [37,40,113,119,171,177] | [37,40,64,99,113,171] | | [37,64,99] | [33,177] |
| Complementary | [11,52,74,126,156] | [34,81,146] | [5,52,81,90,93,126] | [34] | [5,52,93,126,145,156,166] | [57,93] |
| Development | | | | | | |
| Initiation | | | | | | |
| Expansion | [173] | [125,173] | | [115] | [115] | |
| Other | [18] | | | | [18] | |

## 4.1  WHY QUALITATIVE INQUIRY? THE VALUE FOR VOICE UX

Voice is a qualitative medium, and voice interaction takes advantage of its qualities in technology used by people. It should then come as no surprise that most of the reasons behind exploring voice systems qualitatively appear to match the reasons for exploring voice quantitatively [24,139,141]. Yet, qualitative approaches to voice aim for different forms of outcomes and raise new perspectives. Voice characteristics, linguistic utterances and conversation, non-linguistic utterances and vocal sounds, and even singing … these are all non-numerical forms of expression that have qualitative implications for HCI. We summarize the value of qualitative inquiry for voice UX across phenomenological, methodological, and ethical factors in Table 5. As our findings show, qualitative inquiry can reveal as well as deeply illustrate specific features of voice UX. But the voice is only one factor. The eyes-free and hands-free interaction opportunities, for instance, have implications for multi-tasking and people with disabilities, while also raising challenges for interaction design. How can we leverage voice input for storage or word processing, especially when we wish to review and revise our recordings? Voice is ephemeral. Adding supplementary modalities, such as text and visuals, can fix some problems but introduce others. Suddenly the power of an audio-only mode of interaction is disrupted.

*Table 5. Summary of factors revealed by qualitative inquiry for voice UX.*

| Voice UX Factors | Qualitative Factors of Voice UX |
|---|---|
| **Phenomenological** | |
| Experiential | Eyes-free interaction; hands-free interaction; ephemerality of voice input/output; trade-offs with supplementary modalities and multi-modality; longitudinal UX and long-term engagement; richness and complexity of real contexts of use; anthropomorphism; role of the body; diversity in individual responses |
| Attitudinal | Mental models, esp. over time and with changes to the voice or underlying technologies; orientations to and relationships with the system; relative awareness of trust and privacy issues, esp. with voice cloning and voice impersonation |
| Social | Experience of people with disabilities and marginalized populations; multiple users at the same or different times; peripheral users who may unintentionally "listen in" |
| Vocal | Vocalics and voice characteristics; utterances and conversation; non-linguistic utterances; vocalizations and vocal sounds |
| **Methodological** | Mixed methods; overreliance on a few approaches to qualitative inquiry and data collection; limited tasks and activities |
| **Ethical** | Trust and over-trust; privacy and security, esp. data storage, transcripts, who has access or can "overhear"; voice cloning and impersonation; task designed as one-way prescription; decision-making around changing technologies that people have become familiar with and reliant on; disruptions caused by the presence or mismatch between voice and body; handling disclosures of sensitive information |

As the corpus of work reviewed here shows, qualitative research often takes place within real-world contexts over long spans of time, unlike quantitative work. While this can lead to more ecologically valid findings [20], it also sparks new lines of inquiry related to time, notably:

- how expectations for and mental models of the device evolve
- whether and how time of day or day of the week matters [149]
- whether a relationship with the system comes into being, how, and what its nature is over time

What if the voice system changes—what if a new version comes out? If the default voice is made more realistic or is re-voiced, does it become more acceptable because anthropomorphism is ideal? Or will it be rejected, just as Sir. Stephen Hawking rejected an "updated voice" for his voice generator?[11] We can begin to map out not just the quantitative "if" but also the qualitative "why" and "how" to such questions.

Qualitative inquiry was often found deployed hand-in-hand with quantitative approaches: the mixed methods format. Yet, the majority of the work employed the qualitative side in similar and potentially limited ways:

- Most qualitative approaches served as complements to the main quantitative strand
- Most employed interviews as a technique of data collection, echoing the notion of a practical, positivist-leaning semi-structured take to qualitative inquiry [8]

A few others, however, approached qualitative inquiry in other ways that could be accelerated:

- Taking on expansive approaches, where the qualitative component acts as a provocation or corrective measure
- Taking advantage of the vocal format, particularly conversational scripts that emerge during interaction

Indeed, deriving new intersections between qualitative and quantitative forms of data and metadata may be necessary to understand the relationship between the voice and other matters of embodiment. For instance, we are now able, with some degree of accuracy, to mock up uncanny visions of the face of the voice [109][12]. Machine learning is making these "vocal deepfakes" possible: virtual clones of voices that can trick human ears by reproducing the qualitative features of a person's voice [15]. While several works in this review considered matters of security and privacy, highlighting the notion of unintentional third-party use, e.g., overhearing, transcript access, sharing phones, few explored the user identification front. Are people aware that their voice could be used to reveal who they are? This is not a matter of deriving identity from what are largely social constructs. We are living in a world of big data, where we freely offer our personal bits and bytes to the world, often without realizing the implications [84] or preserving control [2]. We can expand the qualitative orchestra, in concert with quantitative methods, for methodological harmony and deeper engagement with such emerging matters of great concern in voice UX.

---

[11] https://www.npr.org/2018/03/27/597390626/an-engineers-quest-to-save-stephen-hawkings-voice

A critical look at the findings reveals some patterns that deserve interrogation, which we highlight as follows:

- Much of the work deployed voice as a means of compliance and prescription through one-way lectures, e.g., [5,42,72,130,145,163,173]. This was especially the case in medical interactive voice response systems. From a technical perspective, it is not hard to imagine why this may be common. Displaying information, including through an audio channel, is far easier than developing an interactive experience, which typically requires the use of advanced techniques like NLP. Yet, this becomes an issue of power and calls into question the descriptor of "interactive."

- Deeper inquiry into the body of the voice is warranted. Robotic bodies, for instance, may not be necessary. As recent work in HRI has shown, a physical form factor may even introduce new challenges, both technically and in terms of optimizing UX, and especially when not designed to match the voice [140,160].

- When it comes to embodiment, phenomenological approaches with a mixed methods component that triangulates quantitative and qualitative data about the body may be ideal (refer to Table 4).

---

[12] https://speech2face.github.io

- Voice systems thought to be "bodiless," such as smart speakers, may not be interpreted that way by everyone, leading to different assumptions about source and who may be "listening in," as some of the work included in this survey found [70,98,145,159,163,168]. This creates a tension when considering the desires of many to have a more personalized experience with voice-based agents. How can we enable personalization and recognition of the user while also ensuring privacy and security?

- Voice interaction with multiple people can introduce new shifts in the perceived embodiment of the device. When merely a purveyor of human-human communication, does the voice system fold into the background? Or does it remain a third party, however sidelined?

- Social identity can play a role. Voice agents that use a real human voice known to the user/s may build trust and help overcome initial barriers to use, as Joshi et al. [72] found. But what happens when the voice breaks down *socially* rather than technically [6,40,70,115]? Also, does experience with the "impersonator" over time go on to influence future experiences with the real deal? The influence may go both ways. Qualitative research can bring these multiple strands of experience together against trust and privacy factors (refer to Table 4).

- People tend to prefer anthropomorphic forms of embodiment, not just in terms of voice but also body and behaviour. This has implications when people treat these voice agents in negative ways ... or when an agent treats the user poorly, as Rehm explored [125], even if it is a technical glitch. Designing agents that are vulnerable, as in Martelaro et al. [93], may lead to more disclosures, but what are the repercussions for the person? Do they realize what they are disclosing and to whom? Are roboticists, researchers, companies, governments, and other stakeholders willing and prepared to handle sensitive information?

Reflexivity is a powerful method of self-discipline and self-discovery often employed in qualitative research [129]. We can take up this method to shift the scope of inquiry beyond the user to other stakeholders who must also grapple with voice UX and highlight ethical issues on the peripheries. This will require drawing from the literatures outside of computer science and engineering. Biology and neuroscience have explored the possible evolutionary emergence of human conversation for the purpose of creating persuasive narratives so that interlocutors can collaborate and develop shared mental models. The selective advantage offered by grammatically complex codes may have driven the shift away from the simpler speech of our primate ancestors [44]. Whether ideas are abstract or concrete, an affective dimension appears to play a role across human languages [170]. Voices can also be used to signal social identity and reinforce social norms [76], a topic that needs greater exploration in voice UX. Efforts such as harmonization and co-creation with intelligent agents that use voice but potentially other auditory modalities, including music and non-vocal sounds, are on the horizon [175]. This is where a breadth of work from the humanities and social sciences can enter technical fields and drive new questions and collaborations [48,148]. Indeed, there is a large base of literature that is qualitative in nature. For example, we may be inspired by new, critical, qualitative analyses on the racialization of voice and sound, reflecting on our tendency to design around and research *acousmatic questions* about machine identity by way of the voices being embedded in agentic and anthropomorphic form factors that shape our reactions and interactions [41]. Future work could craft a research agenda for cross-pollination across different fields of study that have already produced bodies of work on voice, even if not on computer-based interactive voice systems specifically. There is much we can learn here.

## 4.2   OPPORTUNITIES FOR INCREASING RIGOUR AND COMPREHENSIVENESS

Qualitative approaches to understanding voice UX add value to the quantitative work and have the potential to generate special insights and knowledge. Yet, our review unearthed several weaknesses and gaps that should be addressed. We outline the major points of concern, and opportunities, for increasing rigour in qualitative approaches to voice UX:

### 4.2.1   *Report All Demographics.*

The "who" of qualitative work is often as important or more so than the "what." Yet, the "who" was at times underreported or not reported at all. For instance, gender demographics of 512 participants across 15 papers, almost one-quarter of the work (22.7%), was not made available. This is not simply a matter of methodological rigour, but a matter of inclusion. As critical scholarship has pointed out, human variables such as gender, age, and race are not only captured for data analysis or even to understand the phenomenon under study, but also to ensure accountability, fairness, and diversity [49,135]. Reporting on these demographics is also important for comparison and consensus-building, even to understand the extent to which generalizations that are repeated within one population may not hold across others. We recognize that it may not always be appropriate to ask about sensitive demographics. Qualitative work can be intimate, and demographics reporting in studies involving small numbers of individuals could risk unwanted and/or unethical identification. Still, a statement on why certain demographics were not captured would be ideal.

Even if one takes the position that qualitative work is not generalizable, we must consider the case of mixed methods, which makes up a significant portion of the work. Is voice UX research WEIRD, with user populations drawn from Western, Educated, Industrial, Rich, and Democratic nations [58,59,89,97,137]? We cannot answer such questions without sufficient reporting of demographics. Additionally, it is surprising, given recent work in HCI spaces (e.g., [69,151]), that there were no participants who identified outside of the gender binary. Researchers can refer to recent work on how to best capture such multifarious variables in accurate and inclusive ways. Gender is particularly tricky because it is often not defined by or for participants [138,150] and is used interchangeably with sex and sometimes sexuality [65]. Gender as a social construct can shift over time and is culturally sensitive. It can be difficult for some people to answer, as they may be unsure of their gender identity, or it may be risky for them to volunteer that information. Response options for gender should aim to distinguish it from sex and allow for a range of responses that are not forced [150]. Otherwise, we may generate invalid data and cause discomfort or worse in participants.

### 4.2.2 Fully Describe the Technology and Voice(s).

While the basic type of technology, one that uses voice for interaction, was generally well-reported, there were notably gaps when it came to the specifics of the systems under study. For instance, it was not clear in one-third of cases whether a natural, human voice was used or a synthetic one, and to what degree the voice "sounded" synthetic or "real." Almost one-quarter did not report on the TTS used, possibly because the researchers believed that it was implied by the system. It may also have been the voice of the researchers themselves, i.e., Wizard of Oz, which is not always disclosed. Additionally, the intelligence underlying the voice was not reported almost half of the time (43.9%), possibly for similar reasons as above. Yet, what the system can do and how it behaves with the people is the crux of HCI research. If Wizard of Oz or simulated intelligence was used, or if only part of the system was automated, this should be reported.

Investigating ideal interactions is important for designing the future of voice UX. Still, we need to distinguish between what is *possible now* and what is *desired for then.* This will also open up pathways for comparing different voice systems and forms of voice interaction. Notably, not many works compared systems, even commercial platforms like Amazon Alexa and Google Home. Qualitative work may not necessarily allow for comparative research designs or aim for generalizability. Nevertheless, some do, and we should not forget about mixed methods forms of inquiry. Full reporting may allow for more fruitful forms of inquiry or post hoc consensus building in follow-up review work.

### *4.2.3   Clearly State the Qualitative Approach and Epistemological Assumptions.*

Most papers (49 or 74.2%) did not explicitly label or describe the qualitative approach undertaken. This limited our ability to categorize the work for this review, but it also limits our understanding of the methodology and epistemological underpinnings of the researchers. Epistemology, or our position on what knowledge is and how much we can have access to it, is in some sense a matter of personal philosophy. At the same time, our epistemological stance has clear implications for research [29,32], which is a social practice. A common and pertinent challenge in negotiating qualitative and quantitative research, especially in a mixed methods frame [29], is positivism versus everything else. Positivism in research is characterized by systematic processes involving precise and replicable measurement of a phenomenon resulting in a quantifiable outcome to which statistical analyses can be applied and a generalizable outcome obtained. The generalizability of the outcome tends to be the highlight of this approach. Qualitative research, on the other hand, is rarely characterized as positivist. Some argue that qualitative outcomes tend not to be generalizable in practice; others argue that they cannot be generalizable and that is not the aim to begin with; and still others argue for consensus and synthesis when appropriate and possible [32,45,118]. For instance, Polit and Beck describe the goal of most qualitative research as being to "provide a rich, contextualized understanding of human experience through the intensive study of particular cases" [118:1452, our emphasis]. They caution that we must be careful when generalizing within as well as between qualitative and quantitative traditions. Creswell and Poth [32], on the other hand, argue for elucidation and transparency (and some degree of pragmatism). We do not necessarily argue for or against treating qualitative outcomes one way or the other here. Instead, we wish to call attention to the benefits of, as a first step, clearly describing one's position, so that others can make informed choices and fruitful debate premised in stated positions can then ensue.

### *4.2.4 Explore and Synchronize Novel and Underused Qualitative Approaches and Techniques.*

We identified several forms of qualitative and mixed methods inquiry that appear to be underused, despite their potential value for voice UX. Narrative studies, for instance, are based in story–and what is an interaction, especially one that takes place over time, but a story? What is a conversation with a voice-based system but a tale stripped down to its most common form of communicative exchange? The building blocks of narrative approaches have already started to be explored, notably diaries, dialogues and transcripts, design outcomes from workshops and jams, and drawing procedures. We highlight these promising techniques in Figure 3. The outcomes could be used to form a narrative or the techniques themselves could be employed by participants to construct their own narratives about the voice-based systems they use (or may wish to use). Indeed, we may be on the cusp of a narrative revolution in qualitative approaches to voice UX. For instance, Ringfort-Felner et al. [127] explored *design fiction* as a means of mapping out conversations with voice assistants. Similarly, the case study format was all but unused. Yet, a deep dive into a particular device or unusual use case with a range of data collection techniques could be instructive for the present and future of voice UX.

Grounded theory was also underused, but we are not confident that this was truly the case. Many studies relied on common techniques used in grounded theory in terms of data collection and analysis, especially thematic analysis. Yet, they did not refer to grounded theory (in label or reference) and/or did not seem to treat the thematic framework as a theory in the grounded theory sense. This could point to what Blandford [8] has described as a semi-structured approach to qualitative inquiry: harnessing methods from other disciplines in a pragmatic fashion to better understand HCI phenomena in a structured way, or at least in a somewhat structured way. The "semi" nature of this approach disrupts expectations for epistemological stance, deployment, analysis, and triangulation. Importantly, those who take such an approach should faithfully record what they have done and how they have drawn their conclusions so that others "can comprehend the journey from an initial question to a conclusion, assess its validity and generalizability, and build on the research in an informed way" [8:3]. Blandford suggests that this is in some part a naming problem, i.e., there may not be a label for the specific approach used due to deviations from the original methodologies, the pragmatic taking-up of methods from other fields, the mixing of methods and techniques ("bricolage"), and so on. Nevertheless, we urge researchers to clearly mark their work and refer to foundational material and methodological guidance on the methods used whenever possible. Labels can be used even in the case of mixing and deviating from established protocols, e.g., "We used grounded theory with these changes ..." Recognition, acknowledgement, and transparency are key.

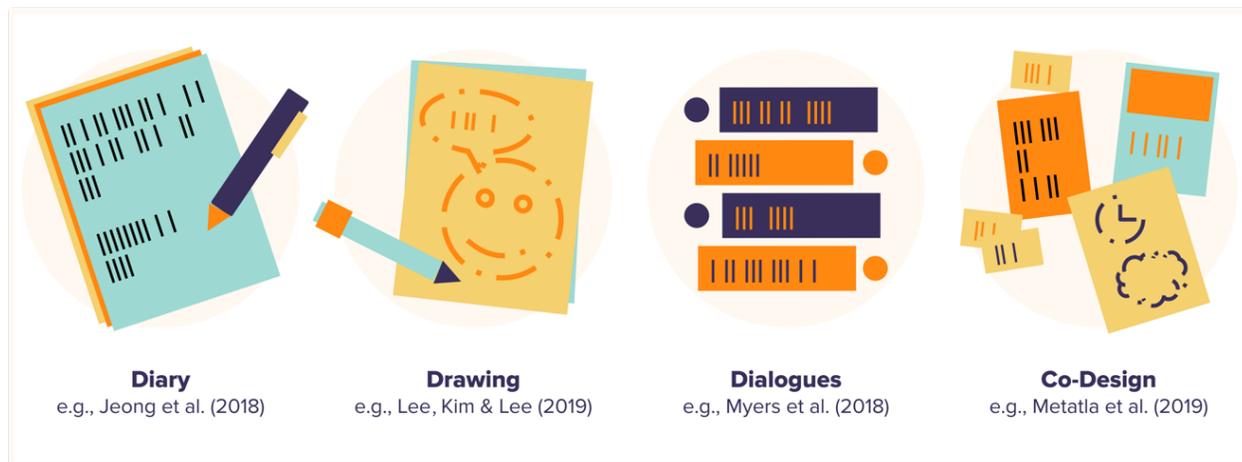

*Figure 3. Four emerging techniques for exploring voice UX qualitatively.*

On the mixed methodology side, two forms of "mixing" were *not* found: development and initiation. The dearth of work taking a development approach may be due to the nascency of voice UX research generally. We may find that, in time, the results of qualitative work may be used to inform the design of future studies, including quantitative measures, instruments, and research questions. The lack of orientations towards initiation in mixing methods is harder to explain. Yet, the goal is a bit counterintuitive: rather than seeking stable patterns, meaning, and consensus, initiation-based approaches aim for provocations and unearthing contradictions that cannot be easily explained–or found.

We highlight these frames for future research. Indeed, there is great opportunity to mix the qualitative and quantitative in new and provocative ways. Where possible, we should aim to use the same tools, and validate these tools, when investigating a particular phenomenon of study or measure of experience. Otherwise, we may not be able to determine if any change is due to our tools or our representation of reality as researchers and/or users. A 10-year retrospective on a single tool could be enlightening.

## 4.3   ASK THE SAME QUESTIONS ABOUT THE SAME NOTIONS? TOWARDS A MIXED METHODS FUTURE

Knowledge creation is a multifaceted, social activity that benefits from a plurality of perspectives and research activities. Qualitative research is often valued for its ability to find something new rather than confirm what may already be known. Indeed, as our synthesis of findings has shown, the qualitative research on voice UX to date has illuminated an array of phenomena primed for further inquiry–whether qualitative, quantitative, or some combination of the two. Moreover, future synthesis on a given phenomenon would benefit from shared foundations. Specifically, we mean theoretical or conceptual bases and research questions. Similar to a recent quantitative survey [139], nearly half of studies did not reference theory or employ a theoretical model (46% versus 40.9%). Qualitative research is often theory-generating rather than theory-confirming, which could explain this. Still, only five studies (7.6%) were explicitly oriented this way, i.e., employed a grounded theory approach. It is possible that more of this work could be used in a theory-generating way based not on the qualitative orientation but rather on data analysis methods used, especially thematic analysis. Follow-up work could trace how the results of such qualitative work has subsequently been used to guide theorizing, confirmatory studies, or further exploration ... and then use it. Employing different methods to explore the questions derived from qualitative inquiry and/or the same questions in a multi-methods manner may be effective for discovery activities but also would make for ease of comparison between experimental and non-experimental survey work at a high level in future. We do not suggest limiting research questions per se, but rather strive for a combination of novelty and stability in forms of inquiry, with a view to present and future comprehensiveness. Finally, we did not have the scope in this paper to survey and compare the qualitative and quantitative literatures. Future work can draw on and update existing open data sets on these literatures and carry out these analyses.

## 5   CONCLUSION

Voice UX represents a modern form of engagement with computer systems inspired by one of humanity's longest standing and perhaps essential, forms of self-expression and communication. Experiences with voice interaction are qualitative, and while quantifiable, these subjective, non-numerical engagements deserve appropriate inquiry. We have mapped out the state of affairs on this topic, demonstrating the *what* and *how* of qualitative approaches to understanding voice UX across a range of computer-based agents, interfaces, environments, and

systems–as well as the *why*. We have highlighted pertinent findings, from human factors especially relevant to voice UX, general patterns in UX across the corpus of research so far, and trajectories for future research and theory-building. Now that we have mapped out the what, why, and how, we encourage voice UX researchers to join us in considering and justifying the *where* and *when* of incorporating qualitative forms of inquiry into our work.

## ACKNOWLEDGMENTS

We thank the editor and reviewers for their service. This work was funded by a Japan Society for the Promotion of Science (JSPS) Grant-in-Aid for Early-Career Scientists (KAKENHI Wakate) (Grant no. 21K18005).